# ANALYSIS AND DEVELOPMENT OF A TURBIVO COMPRESSOR FOR MVR APPLICATIONS

Elias BOULAWZ KSAYER, Denis CLODIC

Center for Energy and Processes, Ecole des Mines de Paris
60, boulevard Saint Michel 75272 Paris Cedex 06, France.
E-mail: elias.boulawz_ksayer@ensmp.fr, denis.clodic@ensmp.fr.

## Abstract

The mechanical vapor recompression is an efficient process to decrease energy consumption of drying processes. In order to use the mechanical vapor recompression (MVR) in residential clothe dryers, the volumetric Turbivo technology is used to design a dry water vapor compressor. The Turbivo volumetric machine is composed mainly of a rotor with one blade, a stator, and a mobile oscillating thrust. The advantages of Turbivo® technology are the absence of contact between rotor and stator as well as the oil-free operation. A model of the Turbivo compressor, including kinematic, dynamic, and thermodynamic analysis is presented. The compressor internal tightness is ensured by a surface treatment of the compressor components. Using the model, a water vapor Turbivo compressor of $12m^3/h$ and compression ratio of 5 has been sized and realized. The compressor prototype will be tested on a dedicated test bench to characterize its volumetric and isentropic efficiencies.

## 1. INTRODUCTION

One third of European homes use clothes dryers, which energy consumption is one of the highest of all basic appliances. Reducing it consumption is at the heart of current sustainable development issue. Typical clothe dryers consume 750 Wh/kg of dry clothes. Dryers with thermodynamic cycle can divide by two this consumption. Dryers with mechanical vapor compression are particularly interesting systems: they use water vapor as the working fluid and, in addition, the dryer operation time is twice shorter than usual clothe dryers due to higher heating capacity of water latent heat. The key point of MVR dryers is the use of a compressor with high efficiency in an open cycle and at high evaporating temperature (from 95°C to 100°C). Besides, the new MVR clothe dryers should remain competitive, so the new compressor should be as cheap and as small as possible.

Different types of compressors exist depending of the operation mode [1]. Types of compressors are divided into two categories: volumetric compressor where compression is made by a volume decrease and dynamic compressor where total pressure is increased by an augmentation of the flow velocity. Dynamic compressors (axial flow and radial flow) are mainly used in multi-stage configuration for high flow rate and high compression power (> 100 kW) [2]. Volumetric compressors are divided in two sub-classes: reciprocating and rotary machines. In general, reciprocating compressor is of piston types. Rotary compressors are smaller than reciprocating compressors. For dry compression, the volumetric compressors show many drawbacks:

- Liquid piston: water leakage.
- Sliding thrust and Scroll: permanent friction.
- Straight lobe: low pressure ratio.
- Helical lobe (screw compressor): lubricant leakage and internal fluid leakage.

Most compressors use a lubricant to decrease mechanical friction and to reduce the internal fluid leakage. For clothe dryers with mechanical vapor compression, the drying cycle is an open cycle; so using a lubricated compressor requests highly efficient lubricant recovery. Therefore, dry volumetric compressors should be investigated as MVR machine.

A totally oil-free compressor has been patented by Northey Technologies LTD [3]. Northey invented the design in which the gas is compressed between precision rotors that contra rotates, closely meshing but never touching (Figure 1). Northey compressor has no metal-to-metal contact so there is neither friction nor corrosion, but this type of compressor is expensive due to the manufacturing precision between the rotating parts.

To design a cheap small dry compressor for MVR dryer, a new compressor technology is used: Turbivo machine. Initially, the Turbivo technology is adapted to volumetric turbine, hence its name: Volumetric Turbine [4].

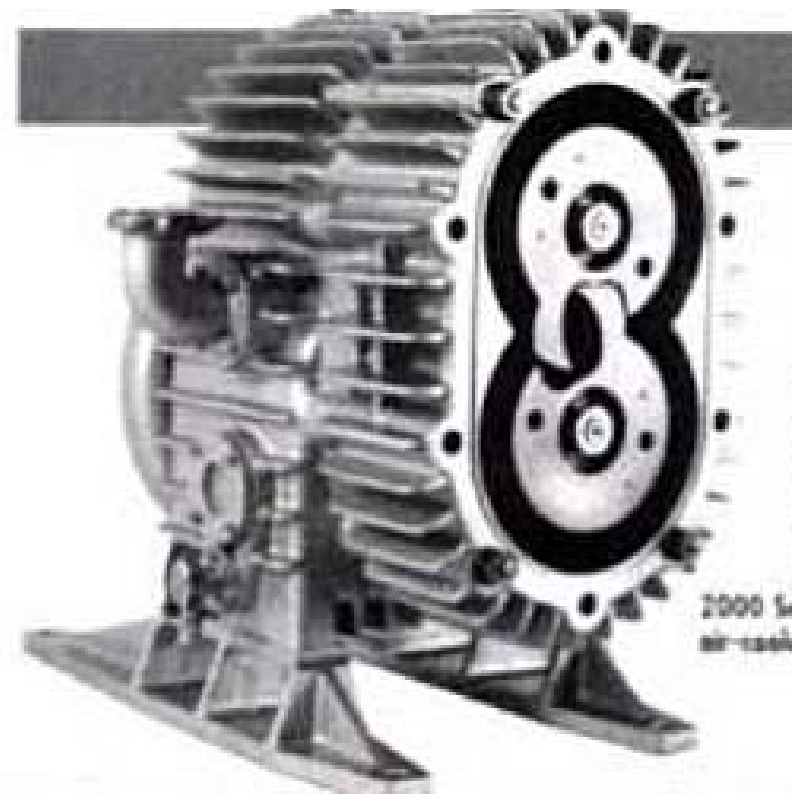
Figure 1 : Northey compressor.

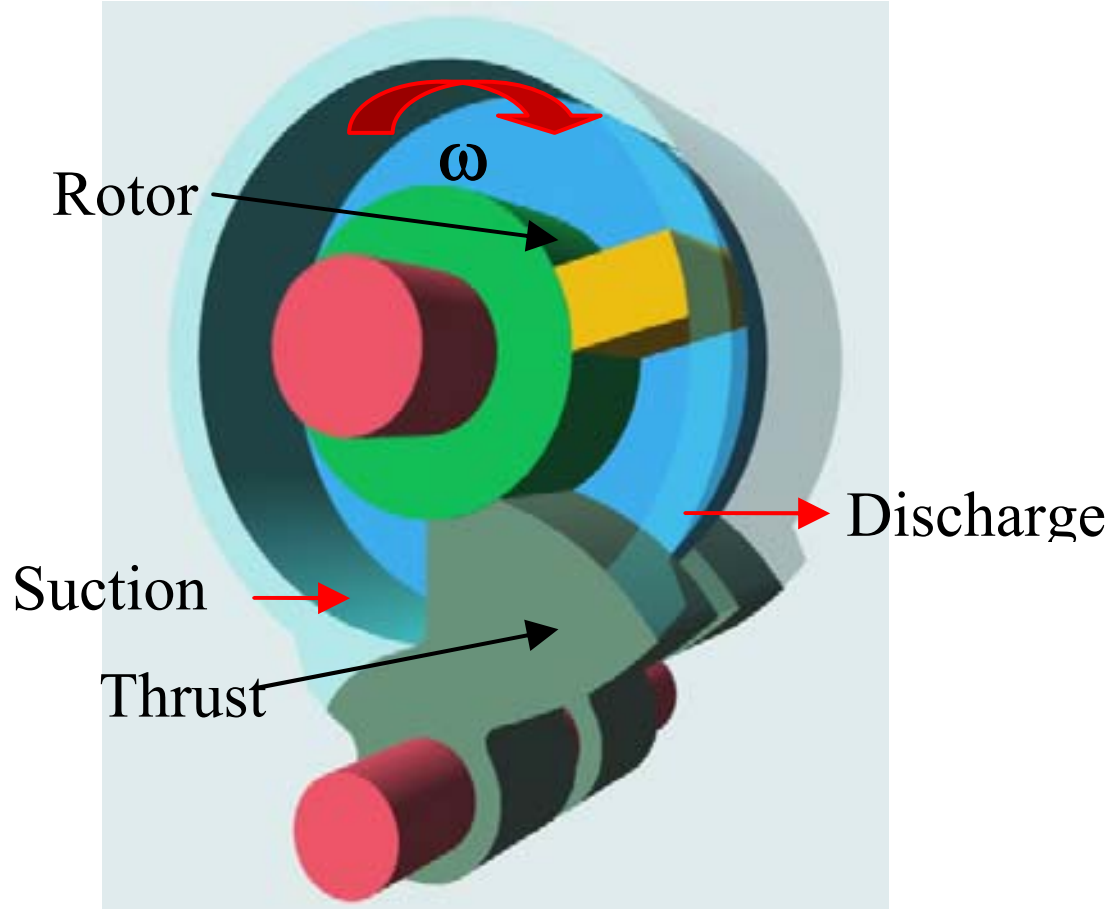

Figure 2: Turbivo initial design.

Like any rotating machine technology, Turbivo can operate as turbine or as compressor. The Turbivo volumetric machine is composed mainly of a rotor with one blade, a stator, and a mobile oscillating thrust (Figure 2). The operating principle of a Turbivo machine is having a rotor with single-blade rotating in a volume closed by oscillating thrust: one side of the rotor sucks the vapor and the other side of the rotor compressed fluid vapor by volume decreasing. A valve is installed at the compressor discharge to ensure the required compression ratio. To ensure a continuous operation of the compressor, the thrust performs an oscillatory motion to pass the rotor blade.

In general, the oscillatory movement of thrust is ensured by a cam mounted on the rotor axis. Hence, there is no contact between moving parts in the compression chamber. Thus, the Turbivo machine is not lubricated and can be used in an open cycle with any type of fluid. To reduce internal leakage of the compressor, the surfaces must have a certain roughness to ensure turbulences in order to limit the leakage flow [5].

Figure 3 shows the layout of the Turbivo compressor. The fluid enters through the suction port located on the left side of the oscillating thrust, and the compressed fluid exits through the discharge port located on the right of the thrust.

## 2. DESIGN MODEL OF TURBIVO COMPRESSOR

The specifications of the MVR dryer compressor are:

- Fluid: water vapor.
- $P_{suction}$ = 100 kPa, $T_{suction}$ = 100°C.
- $P_{discharge}$ = 500 kPa.

To design the Turbivo compressor corresponding to specifications, a model simulating the compressor operation has been developed. This model includes kinematic, dynamic, thermodynamic, thermal analyses, and analysis of the internal leakage.

The described model will be used to design the dryer Turbivo compressor.

### 2.1 Kinematic analysis of TurbiVo compressor

The key point of continuous operation of the TurbiVo compressor is to ensure reliable oscillatory and repetitive motion of the thrust. The oscillatory motion is provided by a cam mounted on the rotor axis. The motion of the TurbiVo compressor thrust (Figure 3) is divided into four phases: low-position phase, rise phase, high-position phase, and return phase.

The thrust low-position phase is determined by the angle $\alpha_1$ (Figure 3.b). The angle $\alpha_1$ determines the operational part of a compressor revolution: from the compression inception to the discharge end. The thrust rotation angle s is nil: $0 < \theta < \alpha_1$, $s = 0$; $\theta$ is the rotor rotating angle.

The thrust rise phase is determined by the angle $\alpha_2$ (Figure 3.c). The angle $\alpha_2$ determines the thrust rise time in order to avoid the contact with the rotor blade: from the beginning of the thrust rise to the high thrust position. The thrust rotation angle s varies between 0 and $\mu$, $\mu$ is the maximal thrust oscillating angle: $\alpha_1 < \theta < \alpha_1 + \alpha_2$, s is function of time t and geometric parameters of the compressor.

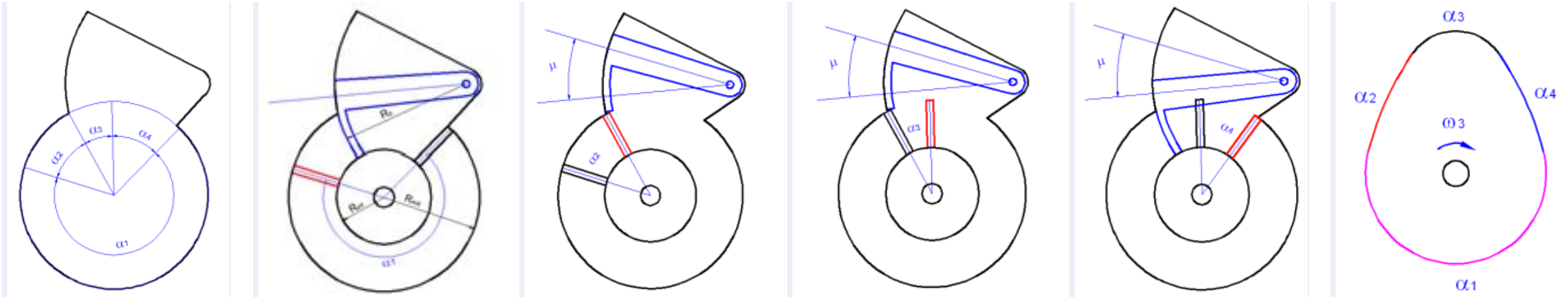

a. Four phases of thrust motion. b. Low-position phase. c. Rise phase. d. High-position phase. e. Return phase. f. Cam ensuring four phases.

Figure 3: Distribution of a compressor revolution on four phases to ensure oscillatory thrust movement.

The thrust high position phase is determined by the angle $\alpha_3$ (Figure 3.d): the thrust should remain in high position for some time to pass the rotor blade. The thrust rotation angle s is equal to $\mu$: $\alpha_1 + \alpha_2 < \theta < \alpha_1 + \alpha_2 + \alpha_3$, $s = \mu$.

The thrust return phase is determined by the angle $\alpha_4$ (Figure 4.e). The angle $\alpha_4$ determines the thrust return time to close the compression chamber: from the high position to the low position of the thrust. The thrust rotation angle varies between $\mu$ and 0: $\alpha_1 + \alpha_2 + \alpha_3 < \theta < \alpha_1 + \alpha_2 + \alpha_3 + \alpha_4$, s is function of time t and geometric parameters of the compressor.

The angle $\mu$ is given by:
$$\mu = 2\, arc \sin\left(\frac{R_{ext} - R_{int}}{2R_b}\right) \qquad (1)$$

$R_{ext}$ is the rotor external radius, $R_{int}$ is the internal rotor radius, and $R_b$ is the thrust radius.

To accomplish the thrust movement, a cam with a follower was used (Figure 4.f). The rise and return phases of the thrust are described by motion laws [6]. Several types of motion laws exist and mainly depend on the cam rotation speed. For the compressor, and since the rotation speed is higher than 1000 RPM, the cycloid motion law (sinusoidal) was adopted. At the beginning and at the end of the movement, the cycloid law presents a nil velocity, a nil acceleration, and a small impulsion. The motion law determines the effort applied on the cam and the follower. The cycloid motion law is given by:
$$s = \mu\left(\frac{t}{T} - \frac{1}{2\pi}\sin\frac{2\pi t}{T}\right) \qquad (2)$$

where T is the period of each thrust phase. By considering that the compressor rotation speed is equal to $\omega$, the periods of the rise and the return phases are given by:
$$T_{rise} = \alpha_2/\omega \qquad (3)$$
$$T_{return} = \alpha_4/\omega \qquad (4)$$

### 2.1.1 Cam description

A disc cam with a follower and a roller was used to accomplish the thrust oscillatory movement of the compressor.

Figure 4 shows the geometric parameters of came-follower-roller system. The follower is fixed to the thrust via an axis passing by point A, the cam is fixed to the rotor via an axis passing by point O. The length $L_1$ is the distance between the two rotation axes, the follower length is $L_2$ and $r_0$ is the came reference radius. The correction angle $\Delta$ is the angle between the polar radius of the follower and the polar radius of the point where the follower center would be if the follower had not made any ascent or descent. The angle $\Delta$ is given by:
$$\Delta = \cos^{-1}\left(\frac{r^2 + {r_0}^2 - 4{L_2}^2\left(\sin\frac{s}{2}\right)^2}{2rr_0}\right) \qquad (5)$$

The sign of $\Delta$ depends on the lengths AD and AO:
$$\begin{cases} AD > AO \Rightarrow \Delta > 0 \\ AD < AO \Rightarrow \Delta < 0 \end{cases}$$

The Ad length is given by: $AD = L_2 \frac{\sin \widehat{ABD}}{\sin \widehat{ADB}}$, $\widehat{ABD} = \frac{\pi - s}{2}$, $\widehat{ADB} = \pi - \widehat{ABD} - (\varphi_0 - s)$. (6)

The pressure angle $\delta$ is the angle between the normal to the cam pitch at the contact point cam-follower and the tangent to the trajectory at this point. The angle $\delta$ will be calculated later in the section on dynamic study of the compressor.

The slope angle $\gamma$ is the angle of the normal to the follower trajectory and the polar radius to the considered point.

The slope angle $\gamma$ is given by:
$$\gamma = \frac{\pi}{2} + \delta - \cos^{-1}\left(\frac{r^2 + {L_2}^2 - {L_1}^2}{2rL_2}\right) \qquad (7)$$

The cam polar radius is given by: $r = \sqrt{{L_1}^2 + {L_2}^2 - 2L_1L_2\cos(\varphi_0 - s)}$ (8)

The angle $\varphi_0$ corresponds to the reference radius $r_0$; so $\varphi_0$ is given by:
$$\varphi_0 = \cos^{-1}\left(\frac{{L_1}^2 + {L_2}^2 - {r_0}^2}{2L_1L_2}\right) \qquad (9)$$

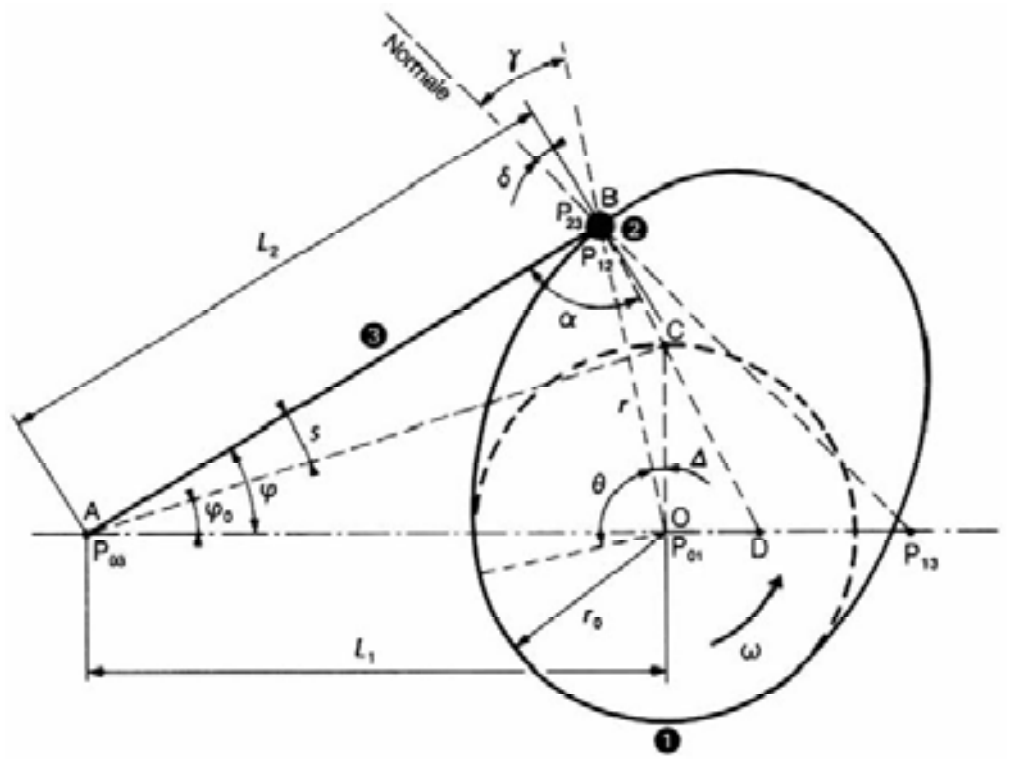

Figure 4: Kinematic scheme of disc cam with a follower and a roller [6].

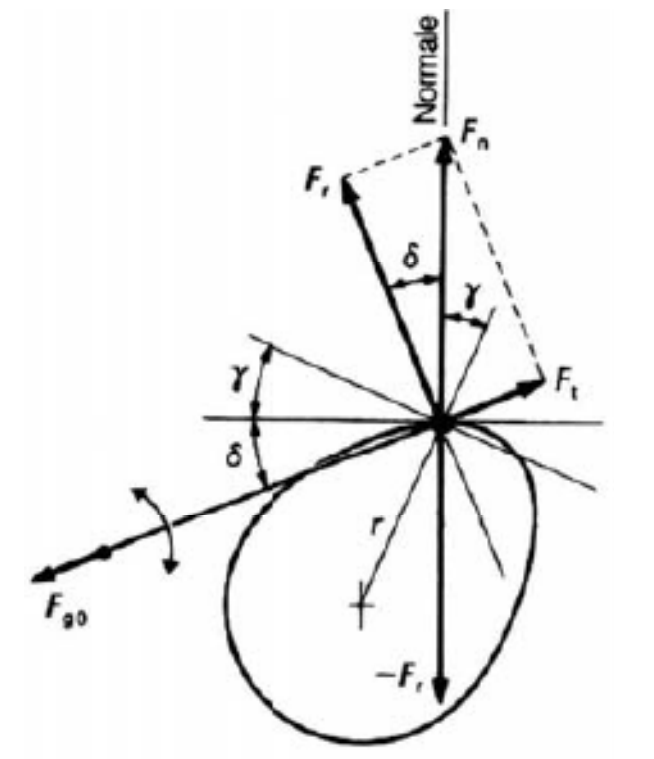

Figure 5: Forces applied on disc cam and follower with a roller [6].

The Cartesian coordinates of the contact point B and its velocity are given by:

$$Position: \begin{cases} x = r\cos(\theta + \Delta) \\ y = r\sin(\theta + \Delta) \end{cases} \qquad (10)$$

$$Velocity: \begin{cases} x' = r'\cos(\theta + \Delta) - r\sin(\theta + \Delta)\,(\theta' + \Delta') \\ y' = r'\sin(\theta + \Delta) + r\cos(\theta + \Delta)\,(\theta' + \Delta') \end{cases} \qquad (11)$$

with the rotating angle $\theta = wt$.
By defining $\alpha_2$ ,$\alpha_3$, $\alpha_4$, $L_1$, $L_2$, and $r_0$, the cam profile is traced. The geometric parameters are determined so as to reduce the efforts mainly at the cam-follower contact point.

### 2.1.2 Determination of the cam curvature radius

The curvature radius of the cam trajectory is used to determine the radius of the follower roller to avoid the interference phenomenon, to ensure permanent contact between the cam and the follower, and to calculate the contact pressures between the cam and follower from the forces acting between them. The curvature radius is given by:

$$\rho_C = \frac{(r^2+V^2)^{3/2}}{r^2+2V^2-ra} \qquad (12)$$

where r is the cam polar radius, V the polar velocity of the contact point, and a the polar acceleration of the contact point.
To ensure a permanent contact between the cam and the follower, the curvature radius must be:

$$\rho_c(t) \geq r_g \;\; \forall t \qquad (13)$$

## 2.2 Dynamic study of Turbivo compressor

The dynamic study of the compressor consists in determining forces on the rotation supports of the cam and the follower, on the contact point between the roller and the cam, and the required torque to ensure the rotation of the compressor. In this section, only the efforts due to the movement of cam-follower system are calculated, and efforts due to compression of the fluid are calculated from the thermodynamic study of the compressor.
In the case of a cam disc with a follower and a roller, the resistive efforts are mainly inertia forces. The resistive inertia torque is given by:

$$C_r = J\ddot{s} \qquad (14)$$

The inertia moment J is the sum of the inertia of the follower, the thrust, and the rotation axis versus the center of the rotation axis, and is given by:

$$J = \int r^2 dm \qquad (15)$$

By referring to Figure 5, the torque applied on the cam is given by:

$$C_c = F_r\, r \tan\delta \cos(\delta + \gamma) \qquad (16)$$

Where $F_r$ is the resistive force at the contact point perpendicular to the follower axis and is given by:

$$F_r = \frac{C_r}{L_2} \qquad (17)$$

The reaction of the axis of the follower $F_{g0}$ is given by: $F_{g0} = F_r \tan\delta$ (18)
The reaction of the cam axis $F_{g0}$ is equal to the normale force at the contact point given by:

$$F_n = F_r / \cos\delta \qquad (19)$$

To evaluate the torque applied on the came, the pressure angle δ should be calculated. The angle δ is calculated by a geometric transformation by referring to the rotating frame (x, y) associated to the cam.

From the velocity coordinates of the contact point (x ', y'), the coordinates of the normale to the cam profile are given by (-y ', x'), so the polar angle ε of the normal is given by:

$$\epsilon = \tan^{-1}\left(\frac{-y'}{x'}\right) \qquad (20)$$

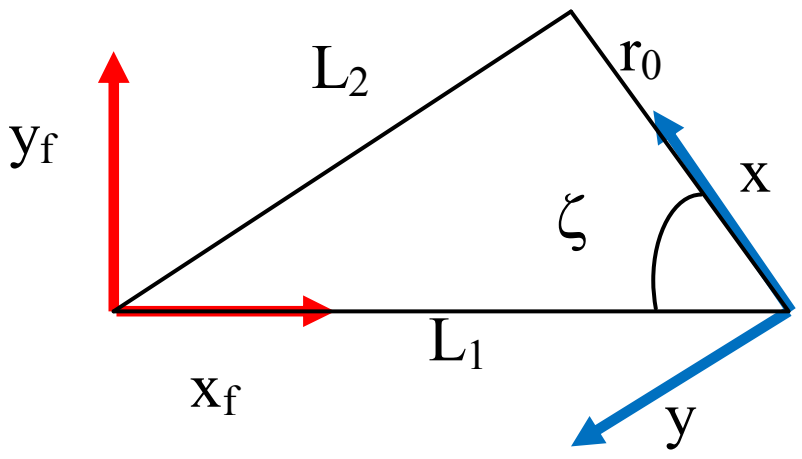

Figure 6: Frames of the follower and the cam.

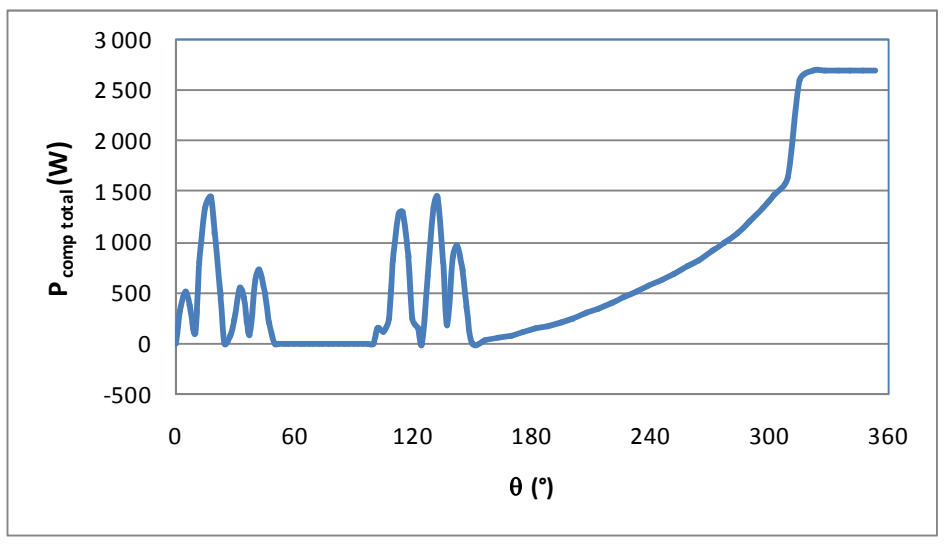

Figure 7: Variation of the total compressor required power with the rotation angle.

To calculate the pressure angle δ, the polar angle of the normal to the follower is evaluated in the frame (x, y) of the cam. To pass from the follower frame ($x_f$, $y_f$) to the cam frame (x, y), a rotation of an angle β and a translation of length L1 should be done. The angle β is given by (Figure 6):

$$\beta = \zeta + \theta - \pi \tag{21}$$

$$\zeta = \sin^{-1}\left(\frac{L_2}{r_0}\sin(\varphi_0 - s)\right) \tag{22}$$

In the frame (x, y), the coordinates of the frame ($x_f$, $y_f$) are given by:

$$\begin{bmatrix} x \\ y \end{bmatrix} = \begin{bmatrix} cos\beta & -\sin\beta \\ \sin\beta & cos\beta \end{bmatrix}\begin{bmatrix} x_f \\ y_f \end{bmatrix} + \begin{bmatrix} L_1 cos(\theta+\zeta) \\ L_1 sin(\theta+\zeta) \end{bmatrix} \tag{23}$$

The polar angle τ of the tangent to the follower in the frame (x, y) is given by:

$$\tau = \beta + \varphi_0 - s + \frac{\pi}{2} \quad if\ s' > 0 \tag{24}$$

$$\tau = \beta + \varphi_0 - s + \frac{\pi}{2} + \pi \quad if\ s' < 0 \tag{25}$$

So, the pressure angle δ will be:

$$\delta = \tau - \varepsilon \tag{26}$$

After the definition of all parameters affecting the cam torque, the cam torque could be expressed as a function of the rotating angle θ.

## 2.3 Thermodynamic study of the Turbivo compressor

The thermodynamic study of the Turbivo compressor consists in evaluating the pressure evolution and in determining efforts during the fluid compression. During the rise and return phases of the thrust, the torque has been calculated and is equal to the cam torque. The compression phase (angle $\alpha_1$) is composed of a compression phase and a discharge phase. The pressure in the compression chamber determines the required applied torque.
Two assumptions may be considered to calculate the pressure evolution during compression:
- Isentropic compression: no change of entropy during the compression.
- Cooled compression: the temperature in the compression chamber is limited to $T_r$ by an injection of a flow of liquid water.

Since the fluid mass in the compression chamber is constant, the density varies with the variation of the compression chamber volume, and the thermodynamic properties of the fluid are calculated as a function of density and entropy, or density and temperature.

## 2.4 Thermal study of the Turbivo compressor

The thermal study of the compressor Turbivo determines the dilatation of compressor parts due to the temperature increase. The part dilatations will be taken into account in determining clearances between moving parts. Once parts are designed, SolidWorks software is used to determine the dilatation values of the parts.

## 2.5 Tightness study of Turbivo compressor

The study of the compressor tightness consists in determining the minimum clearances between fixed and mobile parts to improve the compressor volumetric efficiency, and consequently the overall performance. In the compressor, there are different types of sealing:

- Static Sealing between the stationary parts.
- Dynamic sealing between:
  - the axes of rotation (the rotor and the thrust) and the compressor body,
  - the lateral sides (the rotor and the thrust) and the compressor flanges,
  - the radial faces of the rotor and the thrust,
  - the radial side (the rotor and the thrust) and the compressor stator.

The static sealing is ensured by an O-ring inserted in a groove cut in one side of faces in contact. The dynamic seal between the axes of rotation and the compressor body is provided by rotating lip seals.
For the three remaining cases of the dynamic sealing, it is not possible to use joints, so a permanent leakage rate flows through the clearances of these moving parts. To calculate the leakage flow through a slot, two cases exist: laminar leak and turbulent leak.
For the laminar flow, the leakage rate is given by [7]:

$$Q_m = \frac{\pi d_f^4 \rho}{128 L \mu}(P_e - P_s) \quad (27)$$

For the turbulent flow, the leak flow velocity is given by [7]:

$$P_e - P_s = \rho \frac{V_e^2}{2}\left(1.5 + z(a + 0.5b + \lambda \frac{l}{D_H}\right) \quad (28)$$

## 3. DESIGN OF THE TURBIVO COMPRESSOR PROTOTYPE

Using the model previously described, the Turbivo compressor has been designed and its geometrical parameters have been determined. Figure7 shows the variation of the compressor power with the rotation angle, the compressor power reaches 2700 W during the discharge phase. To ensure permanent contact between the cam and the roller, a disc cam with a double profile groove was used (Figure 8).
To determine dilatations of the compressor moving parts (the rotor and the thrust made in aluminum 6061), simulations have been made using SolidWorks. Figures 9 and 10 show the dilatation of the rotor and the thrust heated to 160 ° C, and submitted to pressure difference of 400 kPa on the rotor blade and the high-pressure side of the thrust; as results, the rotor deformation is less than 50 microns, and the deformation of the thrust is less than 85 microns.
As the rotor has a blade, its gravity center is not the center of the rotation axis, for that to balance the rotor, many holes were drilled (Figure 9) and the holes diametrically opposed to the blade have been filled by carbide metal (density ~ 15). Similarly, to balance the cam, three circular holes have been drilled on the side of the cam small radius (Figure 8).

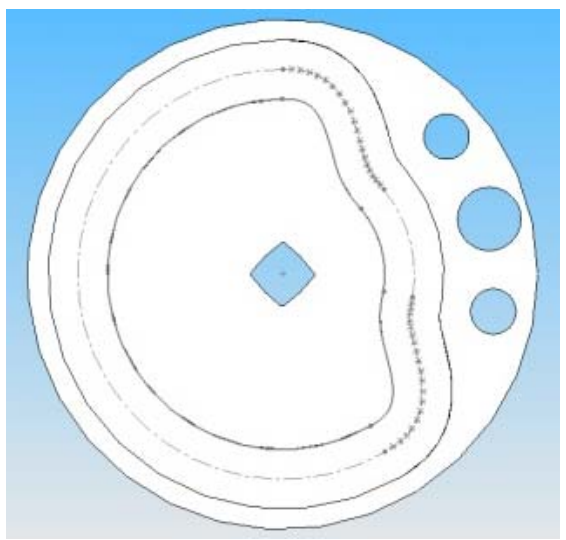

Figure 8: Cam with a double profile groove.

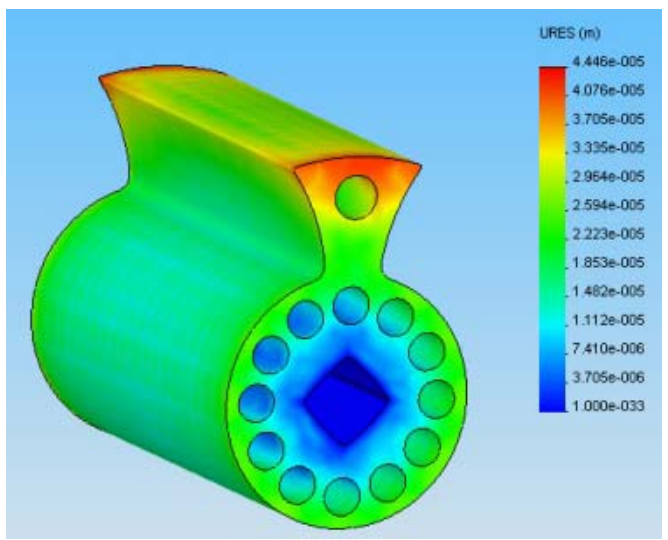

Figure 9: Deformation of the compressor rotor.

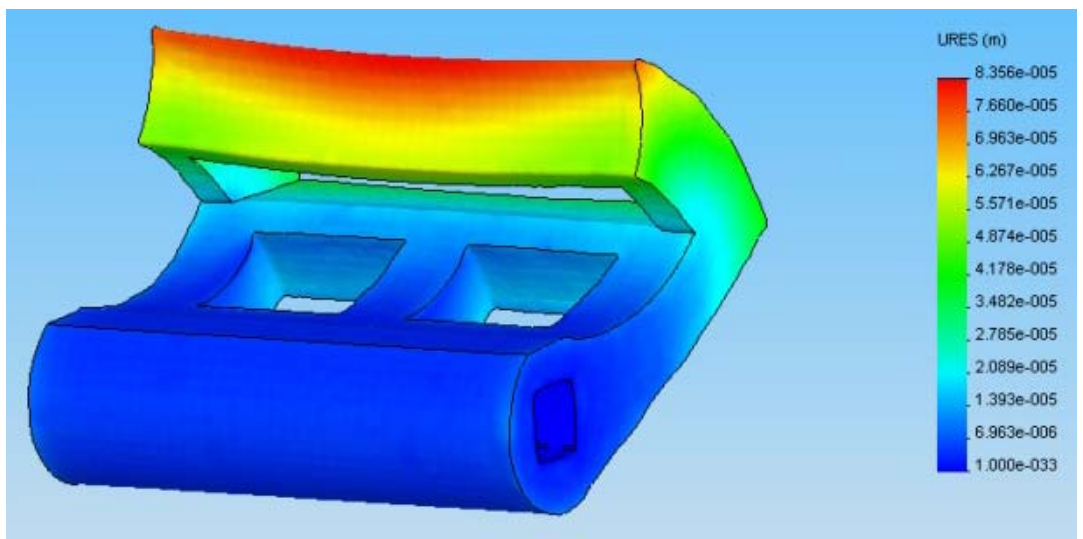

Figure 10: Deformation of the compressor thrust.

Taking into account the realization constraints, a clearance of 0.1 mm was achieved between the rotor, the thrust and the fixed parts of the compressor, and grooves with a depth ***g*** of 0.3 mm were cut on the adjacent surfaces in the direction perpendicular to leak flows: straight lines and arcs (Figure 11). Therefore, internal leakage will reduce the volumetric efficiency of the compressor by about 10%.

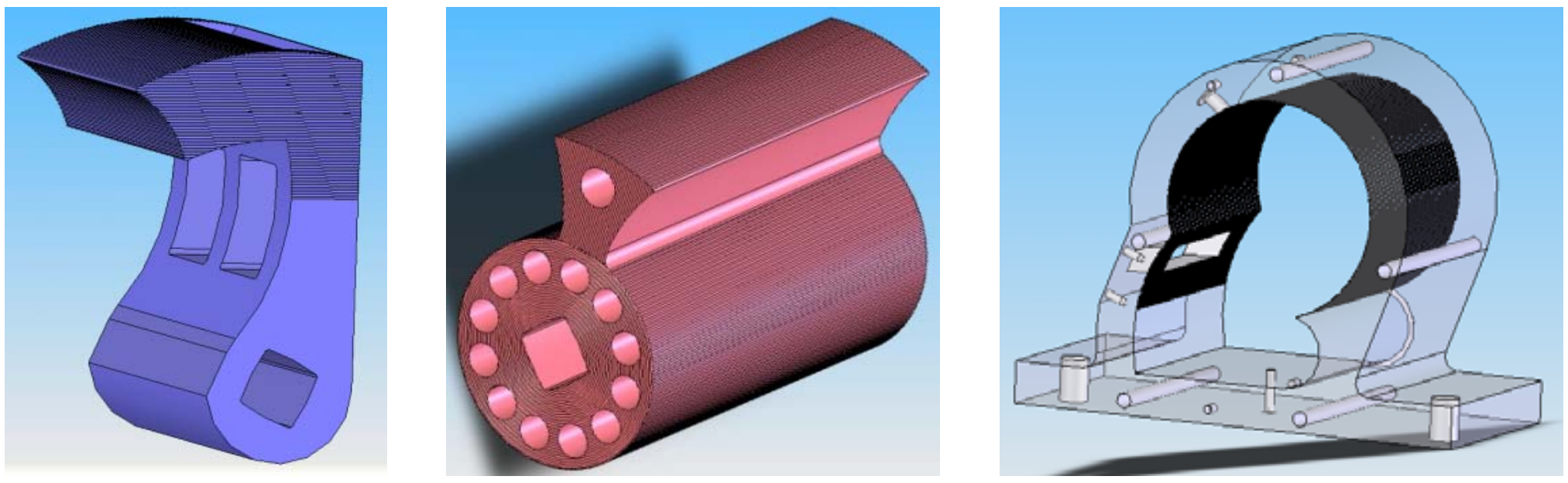

Figure 11: Grooves perpendicular to the leak flow direction on the thrust, the rotor and the stator.

## 4. REALIZATION AND ASSEMBLY OF TURBIVO COMPRESSOR

Once the design of different parts of Turbivo compressor is achieved, the compressor is mounted via SolidWorks to validate its operation and identify any interference. After validation of the virtual operation of the compressor, the compressor assembly drawings were sent for execution. The compressor has 36 components (Figure 12). After the realization of compressor parts, the compressor is assembled and mounted on a test bench to characterize its performances.

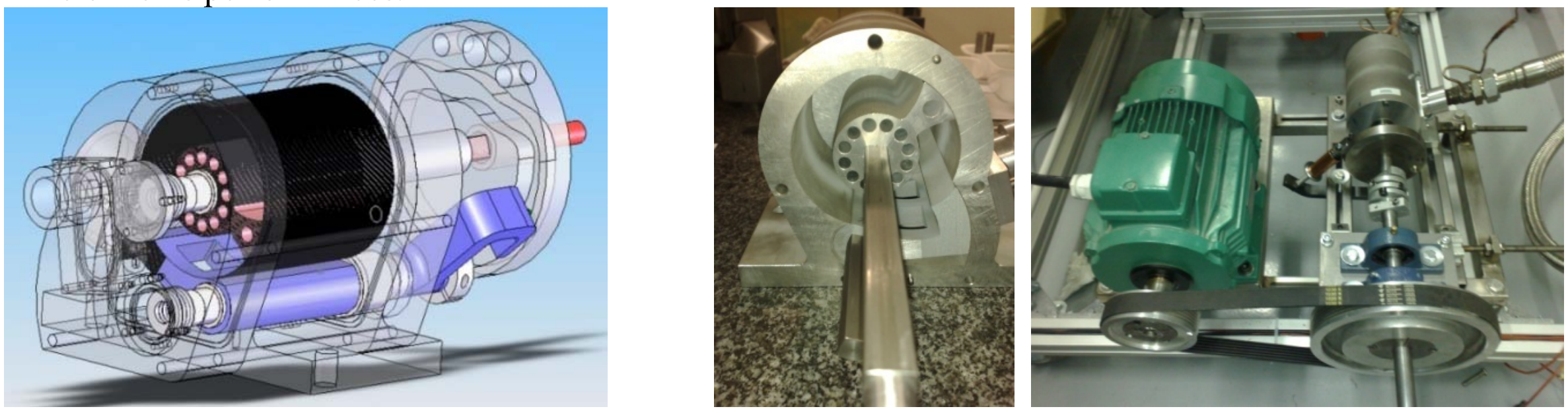

Figure 12: Scheme of the assembled compressor.

## 5. TESTS AND CHARACTERIZATION OF TURBIVO COMPRESSOR

After the compressor assembly, the operation of the cam-follower mechanism was tested first by turning the compressor by hand to be sure of the absence of interference between the moving parts of the compressor. Then, the compressor is mounted on a test bench (Figure 13).

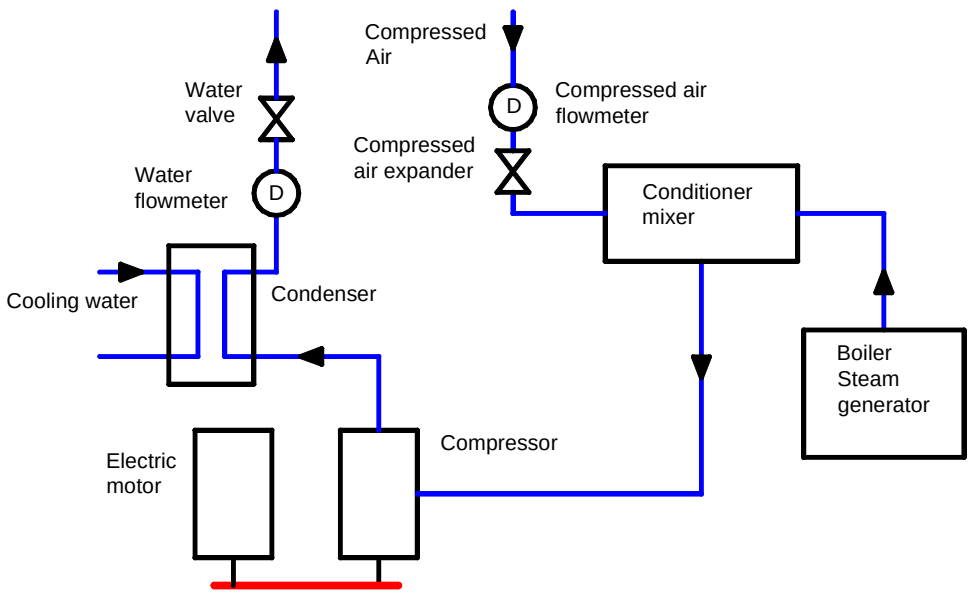

Figure 13: Scheme of the test bench of TurbiVo compressor.

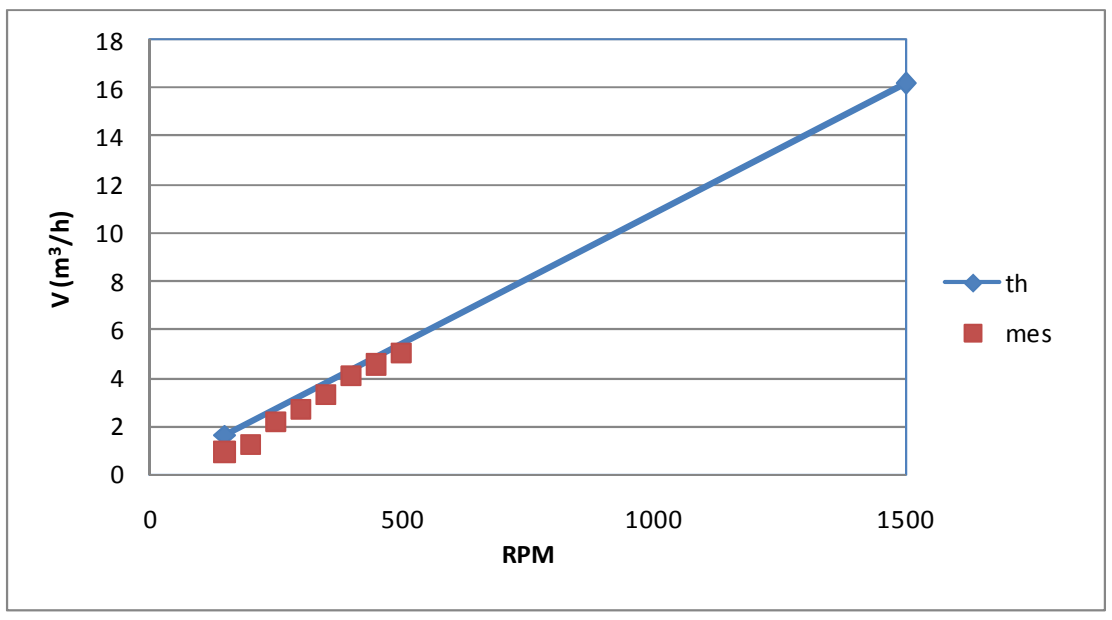

Figure 14: Volume flow rate of the compressor for a pressure ratio of 1.

The purpose of the compressor tests are:

- Validate the operation of the cam-follower mechanism, which ensures the principle of operation of the Turbivo compressor.
- Validate the theoretical compressor volume flow rate for a compression ratio of 1: $P_i = P_o$= 100 kPa.
- Evaluate the volumetric and effective efficiencies of the compressor as a function of the compression ratio up to discharge pressure of 500 kPa with air.
- Evaluate the volumetric and effective efficiencies of the compressor as a function of the compression ratio up to discharge pressure of 500 kPa with water vapor and the behavior of the compressor with the temperature rise from 20°C to 160°C and above.
- Testing the cooling of the compressor by injecting liquid water in the compression chamber.

The compressor rotational speed was varied from 150 RPM to 500 RPM, the compressor volume flow rate for a compression ratio of 1 has been measured, and the results are plotted in Figure 14. The measured flow rate of compressor for a compression ratio of 1 follows the theoretical volume flow rate.

Therefore, two purposes have been validated: the operation of the compressor mechanism and the theoretical volume flow rate. However, by increasing the rotational speed of the compressor to 750 RPM, there have been some wear in the PEEK-Graphite bearing pads, which resulted in a clearance for the rotation axes of the rotor and the thrust, and the compressor has blocked due to contact between the edge of the rotor blade near the flange and the stator. For a clearance of 0.1 mm and a rotor length of 80 mm, the maximum eligible angle of clearance between the axe and the bearing is 0.08°. To develop reliable bearings able to run the compressor at rotation speeds above 500 RPM, metallic non-lubricated bearings will be installed instead of PEEK-Graphite bearings.

## 6. CONCLUSIONS AND FUTURE WORKS

Turbivo technology is a promising technology as dry non-lubricated machine. This technology has been adapted for designing and sizing a dry compressor of steam. This compressor will be used in a Mechanical Vapor Re-compression dryer. The design of a compressor Turbivo has five points to analyze and determine: kinematic study, dynamic study, thermodynamic study, thermal study, and the internal leakage analysis.

A model of Turbivo compressor was developed and used to design a steam compressor with a volume flow rate of 12 $m^3$/h for a compression ratio of 5 and a rotational speed of 1500 RPM.
After determining the characteristics of compressor parts, the compressor was assembled in SolidWorks to validate its kinematic operation. Once the assembly of the compressor has been validated, parts have been realized and the compressor has been assembled.
Then, a test bench dedicated to the characterization of air and steam compressors has been sized and mounted; the compressor was mounted on the test bench.
First test results have validated the operation mechanism of the compressor up to 500 RPM and the theoretical compressor flow; however, the bearing were not well designed to resist to wear, which has blocked the compressor since the clearances between moving parts are 0.1 mm.
So, to ensure the operation of the compressor, supports of the rotation axes have to be strengthened to minimize wear and to ensure continuous operation while keeping the clearance between fixed and moving parts equal to 0.1 mm.
After the design and the installation of strengthened axis supports, the compressor will be mounted on the test rig to characterize its volumetric and effective efficiencies.

**Nomenclature**

| | | | | | |
|---|---|---|---|---|---|
| a | Acceleration of the roller | $m/s^2$ | r | Cam polar radius | m |
| C | Torque | N.m | R | Thrust radius | m |
| $D_H$, $d_f$ | Hydraulic diameter | m | s | Follower angle | rad |
| E | Elasticity module | Pa | t | Time | s |
| $F_r$ | Resistance force | N | T | Valve rise period | s |
| g | Groove depth | m | $R_a$ | Arithmetic roughness | m |
| J | Moment of Inertia | $kg.m^2$ | V | Velocity | m/s |
| l | Groove Length, roller width | m | w | Cam rotation velocity | rad/s |
| L | Flow length | m | x | Abscissa | m |
| N | Rotational speed | RPM | y | Ordinate | m |
| P | Pressure | Pa | z | Number of grooves | adim. |
| $Q_m$ | Mass flow rate | kg/s | | | |
| **Greek letter** | | | **Subscripts** | | |
| $\alpha,\beta,\theta$ | Rotation angle | rad | 0 | at initial time | |
| $\Delta$ | Correction angle | rad | c | cam | |
| $\gamma$ | Slope angle | rad | f, s | follower | |
| $\gamma$ | Poisson coefficients | adim. | i, e | inlet | |
| $\lambda$ | Pressure drop coefficient | adim. | m | | |
| $\delta$ | Pressure angle | rad | o, s | outlet | |
| $\varepsilon$ | Roller tangent angle | rad | int | internal | |
| $\varphi$ | Follower angle | rad | ext | external | |
| $\mu$ | Maximal follower angle | rad | | | |
| $\mu$ | Dynamic viscosity | Pa.s | | | |
| $\rho_c$ | Radius of curvature | m | | | |
| $\rho$ | Density | $kg/m^3$ | | | |
| $\zeta$ | Motion angle | rad | | | |